\documentclass[letterpaper,showpacs,preprintnumbers,superscriptaddress,aps,nofootinbib,preprint,10pt]{revtex4}
\usepackage{amssymb}
\usepackage[centertags]{amsmath}
\usepackage{txfonts}\usepackage{epsfig}
\usepackage{bm}
\usepackage{color}
\usepackage{graphicx,graphics}
\usepackage{multirow}
\usepackage{float}
\usepackage{ulem}
\usepackage{setspace}
\usepackage{slashed}
\usepackage{booktabs}
\allowdisplaybreaks[4]

\begin{document}

\title{Dual Meissner Effect and Quark Confinement Potential in $SU(3)$ Dual QCD Formalism}

\author{Garima Punetha and H.C. Chandola } \footnote{garimapunetha@gmail.com(corresponding author)}

\affiliation{Centre of Advanced Study, Department of Physics, Kumaun University, Nainital-263001, India}

\begin{abstract}
    The mechanism of color confinement has been studied in the framework of SU (3) color gauge theory in terms of abelian fields and
 monopoles extracted by adopting magnetic symmetry. The existence of
 the mechanism of color confinement corresponds to the dual
 Meissner effect caused by monopoles. The two length scales, i.e. the penetration depth and coherence length are defined to demonstrate the scaling
 nature of the QCD vacuum and their ratio defines the Ginzburg-Landau
 parameter indicating the border of type I and type II dual superconductor.
 The existence of these two length scales describes the intrinsic shape of
 the confining flux-tube and is a characteristic of the dual superconductor
 model of confinement in QCD.  As a result, the quark confining potential has been computed and the resulting expression of string tension has been constructed
 in the infrared sector of SU (3) Dual QCD formulation. Moreover, with
 the introduction of dynamical quarks the flux tube breaks and leads to
 the creation of quark anti-quark pairs. Finite temperature quark confining
 potential and the associated string tension has also been extracted which
 demonstrates a considerable reduction in the vicinity of critical temperature showing agreement with the recent lattice studies.

\end{abstract}

\pacs{12.38.-t, 14.80.Hv, 12.38.Aw}

\maketitle

\section{Introduction}\label{S:introduction}

A central issue of elementary particle physics is the description of quark dynamics. One of the remarkable characteristic of quantum chromodynamics (QCD), the asymptotic freedom allows to investigate the quark dynamics at small distances by using perturbation theory. On the otherhand, in its infrared sector, the strong coupling leads to some complicated non-perturbative phenomena such as color confinement. The additional perspective that has been used to describe the quark dynamics at large distances are some phenomenological potential models \cite{Eichten(1978), Martin, Barbashov(1990), Hansenfratz(1978), Adler}, lattice computation technique \cite{Bali(1995)} and some non-perturbative solutions of Schwinger-Dysons equations \cite{Baker(1991)}. A vital illustration of quark dynamics at large distances are the flux tube notion appearing between a static quark and anti-quark. The experimental finding \cite{Regge(1959), Regge(1960), Bugg(2004)} in addition to the lattice QCD simulations \cite{Giacomo(1990), Cardaci(2011), Cea(2012), Cardoso(2013)} are in agreement with the flux tube picture with linear confining potentials and includes the breaking of flux tube owing to production of additional quark anti-quark pairs. To understand the confinement mechanism, an interesting idea \cite{Nambu(1974), Hooft(1979), Mandelstam(1976)} was put forward that describes the quark confinement using the dual version of superconductivity and with such belief, the monopole degrees of freedom and their condensation in the dual superconductor picture of
 Yang-Mills theory \cite{Yang, Kondo(2011), Kondo(2006), Baker(1990)} play the most dominant role in the confinement mechanism and is expected to generate an appropriate quark confining potential. The dynamical evolution of color magnetic monopoles is not demonstrated by the `t Hooft construction, even though there are sufficient lattice evidences \cite{Shiba(1995), Arasaki(1997), Cea(2000), Cea(2001), Giacomo(2000), Carmona(2001), Cea(2004), Alessandro(2010)} to illustrate magnetic condensation in $SU(3)$ QCD vacuum. In the ordinary electric superconductivity tube like structures emerge \cite{Abrikosov(1957)} as a solution to the Ginzburg-Landau equations. Nielsen and Olesen \cite{Nielsen(1973)} found similar solutions in the case of the Abelian Higgs model, where they demonstrate that a vortex solution exists individually on the type I or type II superconductor behavior of the vacuum. The color flux tube made up of chromoelectric field directed towards the line joining of a quark anti-quark pair has also been investigated \cite{Cardaci(2011)} which reveals the characteristics of both superconductor and string models having penetration length and the quantum widening \cite{Allais(2009), Cea(2016)}.  In this direction, one of the popular mechanism to work is the Abelian Projection technique proposed by `t Hooft \cite{Hooft(1981)} that separates the Abelian part and introduces chromomagnetic monopoles by fixing a gauge condition, such as the Maximal Abelian Gauge or Laplacian Abelian Gauge. However, the entire procedure concentrates around choosing to one particular gauge and does not manifest a gauge-invariant confinement mechanism by breaking of gauge symmetry as well as color symmetry.  The existence of monopoles for the mechanism of color confinement has also been investigated  on the lattice by implementing Abelian projection technique elucidating a strong support to the dual superconducting picture of color confinement \cite{Bali(1996), Chernodub(2004), Kondo(1998), Kondo(2011), Kondo(2015)}. In addition, for the better understanding of the phenomenon of confinement several investigations into QCD suggested that a fluctuating flux-tube is formed into a quark and an anti-quark \cite{Bali(1995), Singh(1993), Bissey(2009)} having a finite intrinsic thickness. It is, therefore, desirable to
 develop an approach based on the first principles of QCD, that would provide a clear understanding of the physical picture of QCD vacuum in non-perturbative regime
 and would also allow to perform some analytical calculations in low energy sector of QCD for the study of color confinement phenomena. To resolve the discrepancy, a
gauge-independent description of SU(3) Dual QCD has been proposed which provides a topological
ground to the confinement by imposing an extra magnetic symmetry that leads to a decomposition of the Yang-Mills fields in terms of the electric and magnetic counterparts in a dual symmetric way. The formulation then describes the dual dynamics between the color isocharges and the
topological charges of the non-Abelian gauge symmetry in a viable manner. The SU(3) Dual QCD model is based on the Restricted gauge theory of Cho \cite{Cho(1980)} and is different from the  Dual QCD formalism of Baker, Ball and Zachariasen \cite{Baker(1991)}, since the said formalism is not developed in terms of magnetic symmetry, which manifest the topological structure of the symmetry group in a non-trivial way.

In the present paper, we utilize the magnetic symmetry based SU(3) Dual QCD to discuss its confining structure and the associated confinement potential for the full QCD along-with its thermal response. In section 2, the SU(3) Dual QCD formulation based on magnetic symmetry has been analyzed and its resulting flux tube structure has been investigated for the mechanism of color confinement. In section 3, the non-perturbative gluon propagator has been derived from the
SU(3) Dual QCD Lagrangian in the dynamically broken phase of magnetic symmetry and used to extract the static quark-antiquark potential. The influence of dynamical effects of the light quarks on the quark confining potential due to vacuum polarization has been investigated and used to discuss the polarization effect on the quark confinement potential. The temperature dependence of quark-antiquark potential and influence of dynamical quarks on the quark potential at finite temperature has also been discussed and the numerical results for the quark confining potential have been summarized in the last section.

\section{SU(3) Dual QCD formulation and quark confinement potential}
Based on the non-abelian color gauge group, the non-trivial topological structure plays an essential role in the form of magnetic symmetry to establish the magnetically condensed vacuum necessary for the color confinement in QCD \cite{pandey, hccgp2016, gphcc2016, gphcc2018, gphcc2019}. In the present section, the magnetic symmetry has been applied to $SU(3)$ color gauge group which provide a complete mass spectrum of QCD and guarantees the dual Meissner effect for color confinement. In this context, the magnetic structure of the $SU(3)$ color gauge group may be represented in terms of two internal killing vectors, the first one is the $\lambda_3-$ like octet represented as $\hat{m}$ and the other as the symmetric product presented as $\hat{m}^{'}= \sqrt{3} \hat{m}\ast \hat{m}$ which is $\lambda_8-$ like. In this framework, the gauge potential for the $SU(3)$ color gauge theory may be written in the following form,  
\begin{equation}
{\bf W}_ \mu \, =\,  A_ \mu  \, \hat m + \,  A_ \mu^{'}  \, \hat m^{'}- \frac{1}{g}\,
(\, \hat m \times \partial _ \mu \, \hat m)\, - \frac{1}{g}\,
(\, \hat m^{'} \times \partial _ \mu \, \hat m^{'}),
\label{2.1}
\end{equation}
where 
$
A_{\mu}\equiv\hat m \cdot \bf W_{\mu}$ and $A^{'}_{\mu}\equiv\hat m^{'} \cdot {\bf W_{\mu}}$ 
are the $\lambda_3-$ like and the $\lambda_8-$ like electric component along $\hat{m}$ and $\hat{m}^{'}$ respectively. For the $SU(3)$ color gauge group, the magnetic monopoles emerge as the topological charge of the homotopy group,
$\Pi_{2}(G/H)\rightarrow \Pi_{2}(SU(3)/U(1)\otimes U^{'}(1))$ and to illustrate their existence, let us follow a gauge transformation using the following parametrization written as, 
\begin{equation}
U= exp\biggl[-\beta^{'} (-\frac{1}{2}t_{3}+\frac{1}{2}\sqrt{3}t_{8})\biggr] \times e^{-\alpha t_{n}} exp\biggl[-(\beta-\frac{1}{2}\beta^{'}) t_{3}e^{-\alpha t_{2}}\biggr], 
\label{2.2}
\end{equation} 
so that the magnetic vector $\hat{m}$ may be rotated to a fix time independent direction and may be written in the following form,
\begin{equation}
\hat m = \left( \begin{array}{c} \sin \alpha \cos\frac{\alpha}{2}\cos(\beta-\beta^{'})\\ \sin \alpha \cos\frac{\alpha}{2} \sin(\beta-\beta^{'})\\\frac{1}{4}\cos\alpha(3+\cos \alpha)\\ \sin \alpha \sin\frac{\alpha}{2} \cos \beta\\ \sin \alpha \sin\frac{\alpha}{2} \sin \beta\\-\frac{1}{2}\sin \alpha \cos \alpha \cos \beta^{'}\\-\frac{1}{2}\sin \alpha \cos \alpha \sin \beta^{'}\\\frac{1}{4}\sqrt{3}\cos \alpha (1-\cos \alpha)      \end{array} \right).
\label{2.3}
\end{equation}
This leads to the gauge transformed magnetic potential in the following form,
\begin{equation}
{\bf W_{\mu}} {\buildrel U \over
\longrightarrow} g^{-1}\biggl[\biggl((\partial_{\mu}\beta-\frac{1}{2}\partial_{\mu}\beta^{'})\cos\alpha\biggr)\hat \xi_{3} + \frac{1}{2}\sqrt{3}(\partial_{\mu}\beta^{'}\cos\alpha)\hat \xi_{8}\biggr],
\label{2.4}
\end{equation}
where $\hat{m}$ and $\hat{m}^{'}$ transforms into the space-time independent $\xi_3$ and $\xi_8$ components with the electric potential $A_{\mu}$ and $A_{\mu}^{'}$ in the following form,
\begin{equation}
A_{\mu}=-\frac{1}{2g}\sin^{2}\alpha\partial_{\mu}\beta^{'}, \,\,\, A_{\mu}^{'}=0.
\label{2.5}
\end{equation}
The corresponding field strength then takes the following form given below, 
$$
{\bf G_{\mu\nu}}{\buildrel U \over
\longrightarrow} -g^{-1}\biggl[\sin\alpha\biggl((\partial_{\mu}\alpha \partial_{\nu}\beta -\partial_{\nu}\alpha\partial_{\mu}\beta)-\frac{1}{2}(\partial_{\mu}\alpha \partial_{\nu}\beta^{'} -\partial_{\nu}\alpha\partial_{\mu}\beta^{'})\biggr)\hat m 
$$
\begin{equation}
+\frac{1}{2}\sqrt{3}\sin\alpha (\partial_{\mu}\alpha \partial_{\nu}\beta^{'} -\partial_{\nu}\alpha\partial_{\mu}\beta^{'})\hat m^{'}\biggr].
\label{2.6}
\end{equation}
The Lagrangian for the $SU(3)$ dual gauge theory, in principle, may be obtained by substituting the monopole field $\phi$ and $\phi^{'}$ created due to the topological singularities of $\hat{m}$ and $\hat{m}^{'}$ and its regular dual magnetic potentials $B_{\mu}^{(d)}$ and $B_{\mu}^{'(d)}$ in the following form, 
$$
\pounds_{SU(3)}^{(d)}= -\frac{1}{4}F_{\mu\nu}^{2}-\frac{1}{4}F_{\mu\nu}^{'\,2}-\frac{1}{4}B_{\mu\nu}^{2}-\frac{1}{4}B_{\mu\nu}^{'\,2}+ \bar{\psi_{r}}\gamma^{\mu}[i\partial_{\mu}+\frac{1}{2}g(A_{\mu}+B_{\mu})+\frac{1}{2\sqrt{3}}g(A^{'}_{\mu}+B^{'}_{\mu})]\psi_{r}
$$
$$
\,\,\,+\bar{\psi_{b}}\gamma^{\mu}[i\partial_{\mu}+\frac{1}{2}g(A_{\mu}+B_{\mu})+\frac{1}{2\sqrt{3}}g(A^{'}_{\mu}+B^{'}_{\mu})]\psi_{b}+ \bar{\psi_{y}}\gamma^{\mu}[i\partial_{\mu}-\frac{1}{\sqrt{3}}g(A^{'}_{\mu}+B^{'}_{\mu})]\psi_{y}
$$
$$
|(\partial_{\mu}+i\frac{4\pi}{g}(A_{\mu}^{(d)}+B_{\mu}^{(d)})\phi|^{2}+|(\partial_{\mu}+i\frac{4\pi\sqrt{(3)}}{g}(A^{'\,(d)}_{\mu}+B^{'\,(d)}_{\mu})\phi^{'}|^{2}
$$
\begin{equation}
 - m_{0}(\bar{\psi_{r}}\psi_{r}+\bar{\psi_{b}}\psi_{b}+\bar{\psi_{y}}\psi_{y}) -V,
\label{2.7}
\end{equation}
where $F_{\mu\nu}$, $F^{'}_{\mu\nu}$, $B_{\mu\nu}$, $B^{'}_{\mu\nu}$ are the electric and magnetic field strengths corresponding to the potentials $A_{\mu}$, $A^{'}_{\mu}$, $B_{\mu}^{(d)}$, $B_{\mu}^{'(d)}$ respectively, and the quark triplet is represented by $\psi_{r}$, $\psi_{b}$, $\psi_{y}$. The effective potential responsible for the dynamical breaking of the magnetic symmetry with its form reliable in the phase transition study of the SU(3) Dual QCD vacuum given by,
\begin{equation}
V=\frac{48\pi^{2}}{g^{4}}\lambda(\phi^{\ast}\phi-\phi_{0}^{2})^{2}+ \frac{432\pi^{2}}{g^{4}}\lambda^{'}(\phi^{' \ast}\phi^{'}-\phi^{'\,2}_{0})^{2}.
\label{2.8}
 \end{equation}
where $\phi_{0}$ and $\phi^{'}_{0}$ are the vacuum expectation value of fields $\phi$ and $\phi^{'}$.

\begin{figure}[t]
\centering 
\includegraphics[width=5cm]{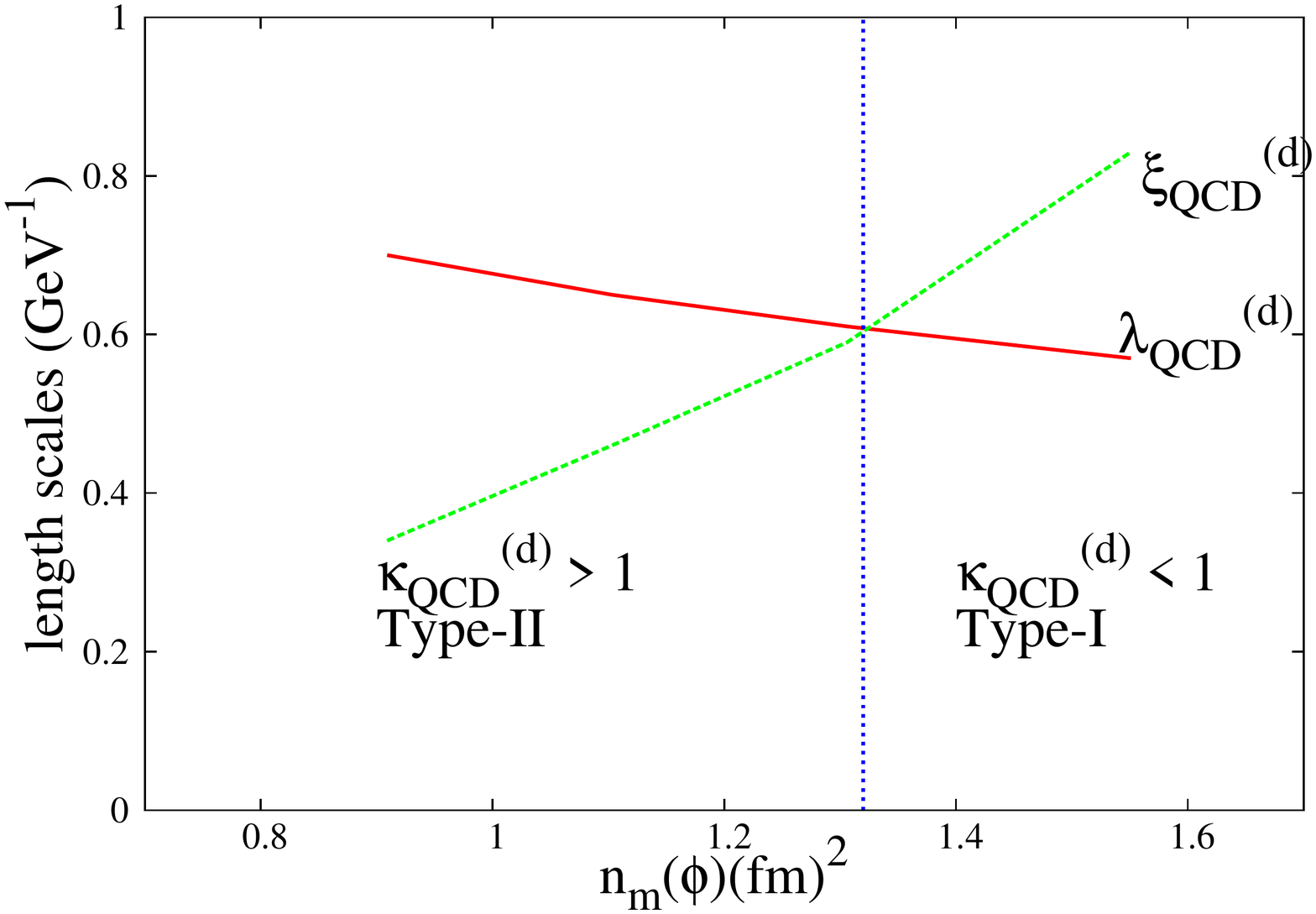}
\includegraphics[width=5cm]{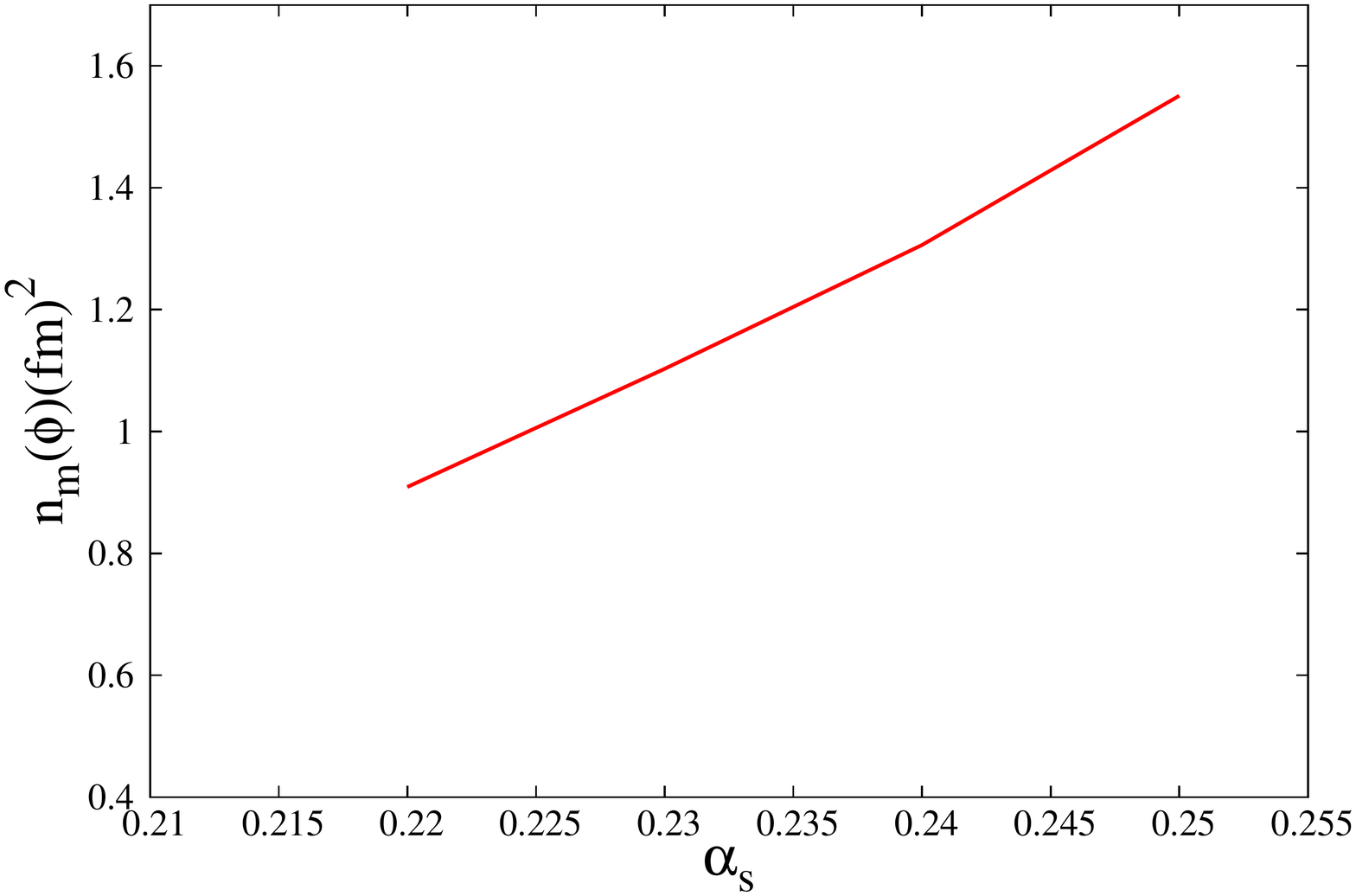}

\caption{(Color online) (a) The length scales as a function of monopole density and
(b) monopole density with coupling in $SU (3)$ Dual QCD vaccum.}
\label{Figure1}
\end{figure}

A crucial insight of the magnetic condensation of QCD vacuum esteems the generation of four magnetic glueballs, two scalars $m_{\phi}$, $m^{'}_{\phi}$ and two vectors $m_{B}$, $m^{'}_{B}$ respectively. However, the existence of the phenomenon of color confinement is incomplete without the incorporation of a residual symmetry, known as color reflection invariance. In order to establish that the physical vacuum is made of color singlets, the property of color reflection invariance is implemented. It tells that the two mass modes $m_{\phi}$ and $m_{\phi}^{'}$ ($m_{B}$ and $m_{B}^{'}$) are the same mode $\bar m_{\phi}$ ($\bar m_{B}$). The estimation of masses has been done in \cite{ChoPRL} and have been presented in table \ref{Table1}. 
\begin{table}
\centering
\caption{The masses of vector and scalar glueball as functions of coupling in $SU(3)$ Dual QCD vacuum.}\begin{ruledtabular}
\begin{tabular}{l c c c c c p{1cm}}

$\lambda$ & $\alpha_s$ & $\bar m_{\phi}(GeV)$ & $\bar m_{B}(GeV)$ & $\lambda_{QCD}^{(d)}(GeV^{-1})$ & $\xi_{QCD}^{(d)}(GeV^{-1})$& $\kappa_{QCD}^{(d)}$\\
\hline
$\frac{1}{4}$ & 0.25 & 1.21 & 1.74 &  0.57 & 0.83  & 0.69 \\

 $\frac{1}{2}$  & 0.24 & 1.68  & 1.63 &  0.61 & 0.59 & 0.99\\

1  & 0.23 & 2.16 & 1.53 & 0.65 & 0.46 &  1.42\\

2   & 0.22 & 2.89 & 1.42 & 0.70  & 0.34 &  2.05\\

\end{tabular}
\end{ruledtabular}
\label{Table1}
\end{table}
The penetration depth ($\lambda_{QCD}^{(d)}$) and coherence length ($\xi_{QCD}^{(d)}$) are related with the vector and scalar glueball masses in the following manner, $\bar m_B = (\lambda^{(d)}_{QCD})^{-1}$ and $\bar m_{\phi} = (\xi_{QCD}^{(d)} )^{-1}$ . The ratio of the characteristic length scales given by, $\kappa_{QCD}^{(d)} =\frac{\lambda_{QCD}^{(d)}}{\xi_{QCD}^{(d)}}$ defines the Ginzburg-Landau parameter. The Dual QCD vacuum behaves as a type- I superconducting vacuum for regions where $\kappa^{(d)}_{QCD} < 1$ whereas it switches to a type-II superconducting behavior when $\kappa^{(d)}_{QCD} > 1$. In the magnetically condensed QCD vacuum, the parameter specifying the confinement mechanism of $SU(3)$ Dual QCD vacuum
is closely related to the density of the condensed monopoles ($n_m (\phi)$) in terms of the
complex scalar field ($\phi$), $n_m = |\phi|^2 = \phi_{0}^2$ . The variation of characteristic length scales with the density of the condensed monopoles in $SU (3)$ Dual QCD vacuum has been depicted in Fig.~\ref{Figure1}a which, in fact, predicts a typical change in QCD vacuum phase from type-I to type-II superconducting state.  The $SU(3)$ Dual QCD vacuum which confines the color electric flux in the form of tubes or filaments, infact, involves a coherent state formed by the condensation of monopole pairs. Consequently, the $SU (3)$ Dual QCD vacuum show type-II and type-I superconducting behavior in relatively weak and strong coupling or density sector respectively. Such a coherence among monopole pairs is completely
lost at extremely low values of the strong coupling constant where the QCD vacuum
comes out from magnetically condensed state and the magnetic symmetry is restored
back which pushes the QCD vacuum into a purely perturbative phase. The drop in
condensed monopole density in the regions of weaker couplings has been shown in Fig.~\ref{Figure1}b, in fact, leads to the
unrealistically thin flux tubes because it corresponds to the high energy sector where
the non-perturbative features of Dual QCD start disappearing in a quite significant
manner.
\section{The quark confining potential in SU(3) Dual QCD formalism}
For the field theoretical description of the $SU(3)$ Dual QCD vacuum, it is necessary to build a singularity free system containing electric and magnetic potentials. The monopoles are described in terms of magnetic potential with a spacelike potential while the isocharges are described in terms of electric potential with a timelike potential. These asymmetries causes singularities. Therefore, using the Lagrangian ``Eq.(\ref{2.7})'', and incorporating Zwanziger's formalism, \cite{Zwanziger} we formulate a local Lagrangian that contains electric and magnetic charges without unphysical singularities. Using Zwanziger formalism, the electric and magnetic fields may be written in the following form,
$$
\mathcal{G}=(\partial\wedge {\bf A})-(n\cdot\partial)^{-1}(n\wedge k)^{(d)},
$$
\begin{equation}
\mathcal{G}^{(d)}=(\partial\wedge {\bf B}^{(d)})+(n\cdot\partial)^{-1}(n\wedge j)^{(d)},
\label{3.1}
\end{equation}
where ${\bf A}= ({\bf A}_3, {\bf A}_8)$, ${\bf B}^{(d)}= ({\bf B}^{(d)}_3, {\bf B}^{(d)}_8)$. The action obtained after Zwanziger formalism depends on the spinor fields for quarks ($\psi$), scalar fields for monopoles ($\phi$), regular time-like electric ($A_{\mu}$) and magnetic ($B_{\mu}^{(d)}$) potential and a fixed space-like four vector $n_{\mu}$ and may be expressed in the following form, 
$$
S^{(d)}_{SU(3)}= \int d^4x \biggl(-\frac {1}{2n^{2}}[n\cdot(\partial\wedge {\bf A}_3)]^{\nu}[n\cdot(\partial\wedge {\bf B}_3^{(d)})^{d}]_{\nu}+\frac {1}{2n^{2}}[n\cdot(\partial\wedge {\bf B}_3^{(d)})]^{\nu}[n\cdot(\partial\wedge {\bf A}_3)^{d}]_{\nu}
$$
$$
-\frac {1}{2n^{2}}[n\cdot(\partial\wedge {\bf A}_8^{'})]^{\nu}[n\cdot(\partial\wedge {\bf B}_8^{'\,(d)})^{d}]_{\nu}+  \frac {1}{2n^{2}}[n\cdot(\partial\wedge {\bf B}_8^{'\,(d)})]^{\nu} [n\cdot(\partial\wedge {\bf A}_8^{'})^{d}]_{\nu} 
$$
$$
-\frac{1}{2n^{2}}[n\cdot(\partial\wedge {\bf A}_3)]^{2}-\frac{1}{2n^{2}}[n\cdot(\partial\wedge {\bf B}_3^{(d)})]^{2} -\frac{1}{2n^{2}}[n\cdot(\partial\wedge {\bf A}_8^{'})]^{2}-\frac{1}{2n^{2}}[n\cdot(\partial\wedge {\bf B}_8^{'\,(d)})]^{2}
$$
$$
+\bar{\psi}(i\gamma_{\mu}\partial^{\mu}-g\gamma_{\mu}{\bf A_3}^{\mu}\cdot {\bf \lambda}_{3}-m)\psi + \bar{\psi}(i\gamma_{\mu}\partial^{\mu}-g\gamma_{\mu}{\bf A_8}^{'\,\mu}\cdot {\bf \lambda}_{8}-m)\psi 
$$
\begin{equation}
+ {\vert(\partial_{\mu}+i\frac{4\pi}{g}{\bf B}_{3\,\mu}^{(d)})\phi\vert}^{2} + {\vert(\partial_{\mu}+i\frac{4\pi\sqrt{3}}{g}{\bf B^{'\,(d)}}_{8\,\mu})\phi^{'}\vert}^{2}
- V \biggr).
 \label{3.2}
\end{equation}
The first four terms are the interaction between the electric and magnetic potentials ${\bf A}_3$, ${\bf B}_3^{(d)}$ and ${\bf A}^{'}_8$, ${\bf B}_8^{'(d)}$ respectively. The fifth, sixth, seventh and eighth term describes the electric and magnetic potentials ${\bf A}_3$, ${\bf B}_3^{(d)}$, ${\bf A}^{'}_8$ and ${\bf B}_8^{'(d)}$ respectively. The ninth and tenth term represents the quark terms and the eleventh and twelfth term describes the interaction of the gauge field ${\bf B}_3^{(d)}$ and ${\bf B}_8^{'(d)}$ with the scalar monopole field $\phi$ and $\phi^{'}$ respectively.

Choosing the effective potential given by ``Eq.(\ref{2.4})'', the dynamical breaking of magnetic symmetry generally leads to the condensation of magnetic monopoles and impart color confining properties to the $SU(3)$ Dual QCD vacuum and therefore the $SU(3)$ Dual QCD action may be written in the following form,
$$
S^{(d)}_{SU(3)}= \int d^4x \biggl(-\frac {1}{2n^{2}}[n\cdot(\partial\wedge {\bf A}_3)]^{\nu}[n\cdot(\partial\wedge {\bf B}_3^{(d)})^{d}]_{\nu}+\frac {1}{2n^{2}}[n\cdot(\partial\wedge {\bf B}_3^{(d)})]^{\nu}[n\cdot(\partial\wedge {\bf A}_3)^{d}]_{\nu}
$$
$$
 -\frac{1}{2n^{2}}[n\cdot(\partial\wedge {\bf A}_3)]^{2}-\frac{1}{2n^{2}}[n\cdot(\partial\wedge {\bf B}_3^{(d)})]^{2} -\frac {1}{2n^{2}}[n\cdot(\partial\wedge {\bf A}_8^{'})]^{\nu}[n\cdot(\partial\wedge {\bf B}_8^{'\,(d)})^{d}]_{\nu}+
$$
$$
 \frac {1}{2n^{2}}[n\cdot(\partial\wedge {\bf B}_8^{'\,(d)})]^{\nu} [n\cdot(\partial\wedge {\bf A}_8^{'})^{d}]_{\nu} -\frac{1}{2n^{2}}[n\cdot(\partial\wedge {\bf A}_8^{'})]^{2}-\frac{1}{2n^{2}}[n\cdot(\partial\wedge {\bf B}_8^{'\,(d)})]^{2}+
$$
$$
\bar{\psi}(i\gamma_{\mu}\partial^{\mu}-g\gamma_{\mu}{\bf A_3}^{\mu}\cdot{\bf \lambda}_{3}-m)\psi + \bar{\psi}(i\gamma_{\mu}\partial^{\mu}-g\gamma_{\mu}{\bf A_8}^{'\,\mu}\cdot {\bf \lambda}_{8}-m)\psi 
$$
\begin{equation}
+ \frac{1}{2}m_B^{2}({\bf B}_{3\,\mu}^{(d)})^{2} +  \frac{1}{2}m_B^{'\,2}({\bf B^{'}}_{8\,\mu}^{(d)})^{2}\biggr).
 \label{3.3}
\end{equation}

\begin{figure}[t]
\centering 
\includegraphics[width=5cm]{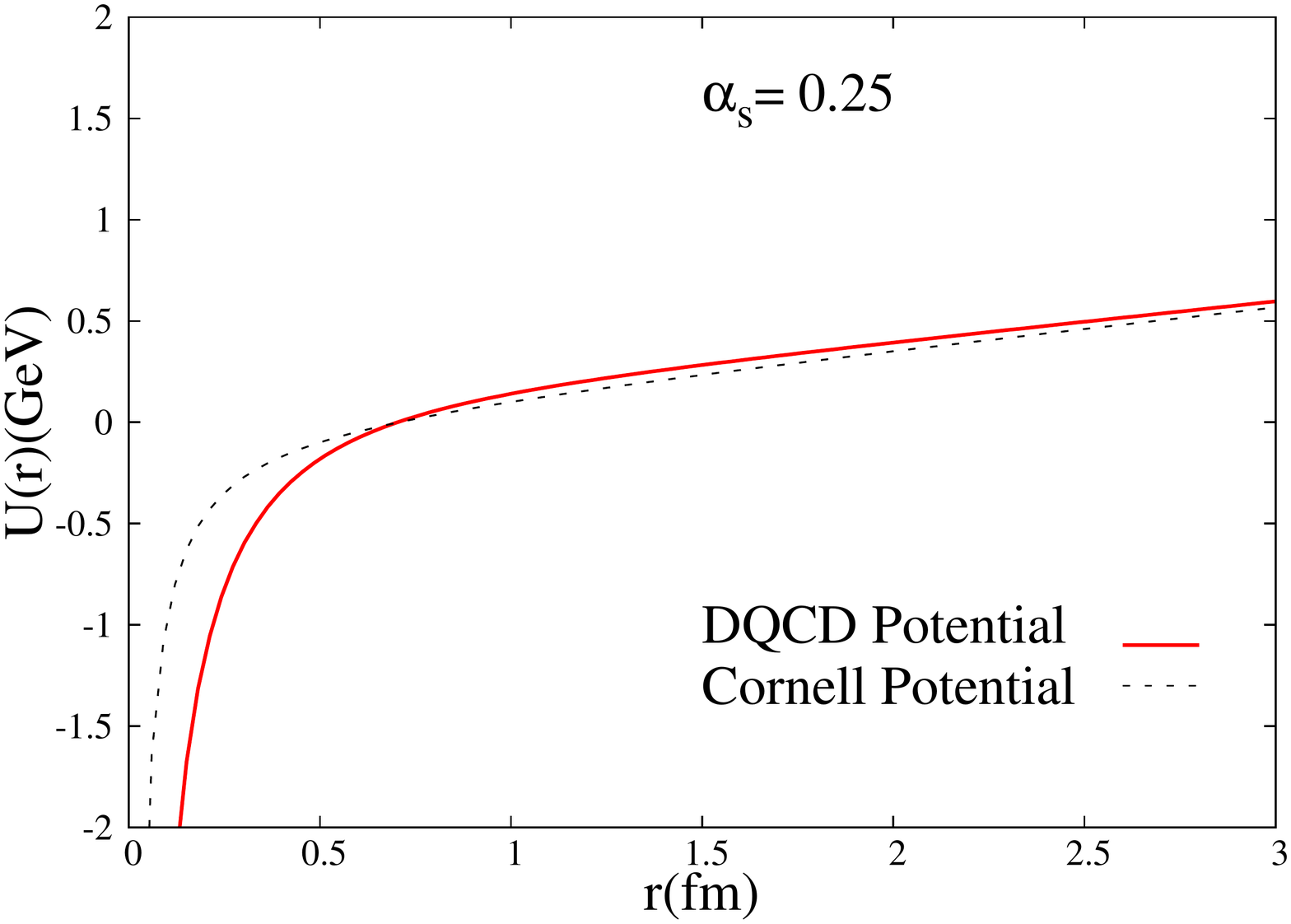}
\includegraphics[width=5cm]{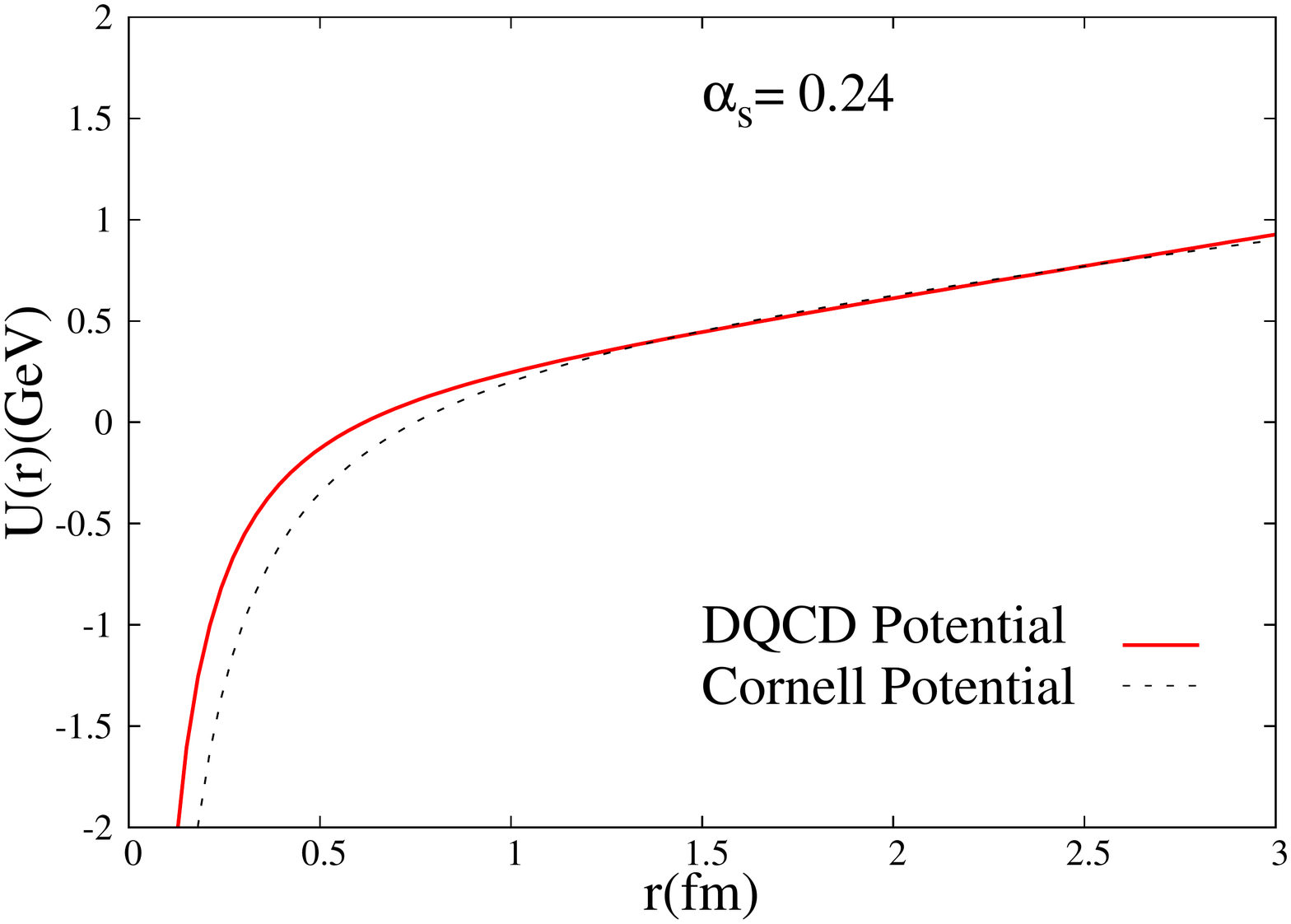}
\caption{(Color online) The quark confining potential for $\alpha_s= 0.25$ and $\alpha_s= 0.24$ coupling in $SU(3)$ Dual QCD vacuum. The dashed curve denotes the Cornell potential.}
\label{Figure2}
\end{figure}

Following the quenched approximation in order to remove the quantum effects of dynamical quarks and eliminating the fields ${\bf A}_{3}$, ${\bf A}^{'}_{8}$, ${\bf B}_{3}^{(d)}$ and  ${\bf B}_{8}^{'(d)}$, the effective action including the quark current is given in the following form,
\begin{equation}
S^{(d)}_{SU(3)}=\int d^4x \biggl[{j^{3}_{\mu}}D_3^{\mu\nu}{j^{3}_{\nu}}+{ j}^{8}_{\mu}D_8^{'\mu\nu}{j^{8}_{\nu}}\biggr],
\label{3.4}
\end{equation}
where $D_3^{\mu\nu}$ and $D_8^{'\,\mu\nu}$ are the propagator of the diagonal gluons,
$$
D_3^{\mu\nu} = -\frac{1}{2}\frac{g^{\mu\nu}}{\partial^{2}+m_{B}^{2}}-\frac{1}{2}\frac{n^{2}}{(n.\partial)^{2}}\biggl(\frac{m_{B}^{2}}{\partial^{2}+m_{B}^{2}}\biggr) \biggl(g^{\mu\nu}-\frac{n^{\mu}n^{\nu}}{n^{2}}\biggr),
$$
$$
D_8^{'\,\mu\nu} = -\frac{1}{2}\frac{g^{\mu\nu}}{\partial^{2}+m_{B}^{'\,2}}-\frac{1}{2}\frac{n^{2}}{(n.\partial)^{2}}\biggl(\frac{m_{B}^{'\,2}}{\partial^{2}+m_{B}^{'\,2}}\biggr)\biggl(g^{\mu\nu}-\frac{n^{\mu}n^{\nu}}{n^{2}}\biggr),
$$
$j_{\mu}^{3}$ and $j_{\nu}^{8}$ are the quark currents with their corresponding Fourier components given in the following form,
$$
j^{3}_{\mu}(k)= Q_{3}g_{\mu 0}2\pi\delta(k_0)(e^{-i{\bf k\cdot b}}-e^{-i{\bf k\cdot a}}),\,\,\,j^{8}_{\mu}(k)= Q_{8}g_{\mu 0}2\pi\delta(k_0)(e^{-i{\bf k\cdot b}}-e^{-i{\bf k\cdot a}}).
$$
The $SU(3)$ Dual QCD action reduces to the following form,
$$
S_{SU(3)}^{(d)}= - Q_3^{2}\int dt\int \frac{d^{3}{\bf k}}{(2\pi)^{3}}\frac{1}{2}(1-e^{i{\bf k\cdot r}})(1-e^{-i{\bf k\cdot r}})\biggl[\frac{1}{{\bf k^{2}}+m_{B}^{2}}+\frac{m_{B}^{2}}{{\bf k^{2}}+m_{B}^{2}}\frac{1}{({\bf n\cdot k})^{2}}\biggr]
$$
\begin{equation}
- Q_8^{2}\int dt\int \frac{d^{3}{\bf k}}{(2\pi)^{3}}\frac{1}{2}(1-e^{i{\bf k\cdot r}})(1-e^{-i{\bf k\cdot r}})\biggl[\frac{1}{{\bf k^{2}}+m_{B}^{'\,2}}+\frac{m_{B}^{'\,2}}{{\bf k^{2}}+m_{B}^{'\,2}}\frac{1}{({\bf n\cdot k})^{2}}\biggr].
\label{3.5}
\end{equation}
The quark confining potential obtained from ``Eq.(\ref{3.5})'' is divided into the following two parts,
$$U(r)= U_{Yukawa}(r)+ U_{Linear}(r),$$
where, $U_{Yukawa}(r)$ gives the Yukawa-type potential,
$$
U_{Yukawa}(r)=- Q_3^{2}\int \frac{d^{3}{\bf k}}{(2\pi)^{3}}\frac{1}{2}(1-e^{i{\bf k\cdot r}})(1-e^{-i{\bf k\cdot r}})\frac{1}{{\bf k^{2}}+m_{B}^{2}}
$$
$$
 - Q_8^{2}\int \frac{d^{3}{\bf k}}{(2\pi)^{3}}\frac{1}{2}(1-e^{i{\bf k\cdot r}})(1-e^{-i{\bf k\cdot r}})\frac{1}{{\bf k^{2}}+m_{B}^{'\,2}},
$$
\begin{equation}
U_{Yukawa}(r)= -\frac{1}{4\pi r}\biggl[Q_3^{2}e^{-m_{B}r} + Q_8^{2}e^{-m_{B}^{'}r}\biggr].
\label{3.6}
\end{equation}
The second term, $U_{Linear}(r)$ is expressed in the following form,
$$
U_{Linear}= - Q_3^{2}\int \frac{d^{3}{\bf k}}{(2\pi)^{3}}\frac{1}{2}(1-e^{i{\bf k\cdot r}})(1-e^{-i{\bf k\cdot r}})\frac{m_{B}^{2}}{{\bf k^{2}}+m_{B}^{2}}\frac{1}{({\bf n\cdot k})^{2}}
$$
\begin{equation}
- Q_8^{2}\int \frac{d^{3}{\bf k}}{(2\pi)^{3}}\frac{1}{2}(1-e^{i{\bf k\cdot r}})(1-e^{-i{\bf k\cdot r}})\frac{m_{B}^{'\,2}}{{\bf k^{2}}+m_{B}^{'\,2}}\frac{1}{({\bf n\cdot k})^{2}},
\label{3.7}
\end{equation}
$$
U_{Linear}=  \frac{Q_3^{2}m_{B}^2}{8\pi^2}\int_{-\infty}^{\infty} \frac{d{\bf k_r}}{\bf k_r^2}\frac{1}{2}(1-e^{i{\bf k\cdot r}})(1-e^{-i{\bf k\cdot r}})\int_{0}^{\infty} d{\bf k_{T}^2}\frac{1}{{\bf k_r^2}+ {\bf k_T^2} + m_{B}^2}
$$
\begin{equation}
+\frac{Q_8^{2}m_{B}^{'2}}{8\pi^2}\int_{-\infty}^{\infty} \frac{d{\bf k_r}}{\bf k_r^2}\frac{1}{2}(1-e^{i{\bf k\cdot r}})(1-e^{-i{\bf k\cdot r}})\int_{0}^{\infty} d{\bf k_{T}^2}\frac{1}{{\bf k_r^2}+ {\bf k_T^2} + m_{B}^{'2}},
\label{3.8}
\end{equation}
$$
U_{Linear}=  \frac{Q_3^{2}m_{B}^2}{8\pi^2}\int_{-\infty}^{\infty} \frac{d{\bf k_r}}{\bf k_r^2}\frac{1}{2}(1-e^{i{\bf k\cdot r}})(1-e^{-i{\bf k\cdot r}})\ln\biggl(\frac{\Lambda^2+ {\bf k_{B}^2}+ m_{B}^2}{{\bf k_r^2} + m_{B}^2}\biggr)
$$
$$
+\frac{Q_8^{2}m_{B}^{'2}}{8\pi^2}\int_{-\infty}^{\infty} \frac{d{\bf k_r}}{\bf k_r^2}\frac{1}{2}(1-e^{i{\bf k\cdot r}})(1-e^{-i{\bf k\cdot r}})\ln\biggl(\frac{\Lambda^{'2}+ {\bf k_{B}^2}+ m_{B}^{'2}}{{\bf k_r^2} + m_{B}^{'2}}\biggr),
$$
\begin{equation}
= \frac{Q_3^{2}m_{B}^{2}}{8\pi}r \ln \biggl(\frac{m_{B}^{2}+\Lambda^2}{m_B^2}\biggr)+ \frac{Q_8^{2}m_{B}^{'2}}{8\pi}r \ln \biggl(\frac{m_{B}^{'2}+\Lambda^{'2}}{m_B^{'2}}\biggr),
\label{3.9}
\end{equation}
where, the cutoff $\Lambda$ and $\Lambda^{'}$ corresponds to the scalar masses $m_{\phi}$ and $m_{\phi}^{'}$ associated with the minimal thickness of the vortex and implementing the property of color reflection invariance, the expression for the quark confining potential in $SU(3)$ Dual QCD vacuum reduces to the following form,
\begin{equation}
U_{SU(3)}(r)= -\frac{\bf Q^{2}}{4\pi}\frac{e^{-{\bar m}_{B}r}}{r}+ \frac{{\bf Q}^{2}{\bar m}_{B}^{2}}{8\pi}r \ln(1+(\kappa^{(d)}_{QCD})^2).
\label{3.10}
\end{equation}
The resulting expression is characterized by a linearly increasing potential at large separations with the proportionality constant given by the string tension ($k_C$) as,
\begin{equation}
k_C = \frac{{\bf Q}^{2}{\bar m}_{B}^{2}}{8\pi}\ln(1+(\kappa^{(d)}_{QCD})^2).
\label{3.11}
\end{equation}  
The variation of the quark confining potential given by ``Eq.(\ref{3.10})'' has been shown in Fig.~\ref{Figure2} for $\alpha_s= 0.25$ and $\alpha_s= 0.24$ coupling. The graphical representation clearly shows a short-range coulomb behavior and a long-range linear rise illuminating the most important features of QCD dynamics, the asymptotic freedom and confinement. The variation is in agreement with the phenomenological Cornell potential \cite{Chung}.

\begin{figure}[t]
\centering 
\includegraphics[width=5cm]{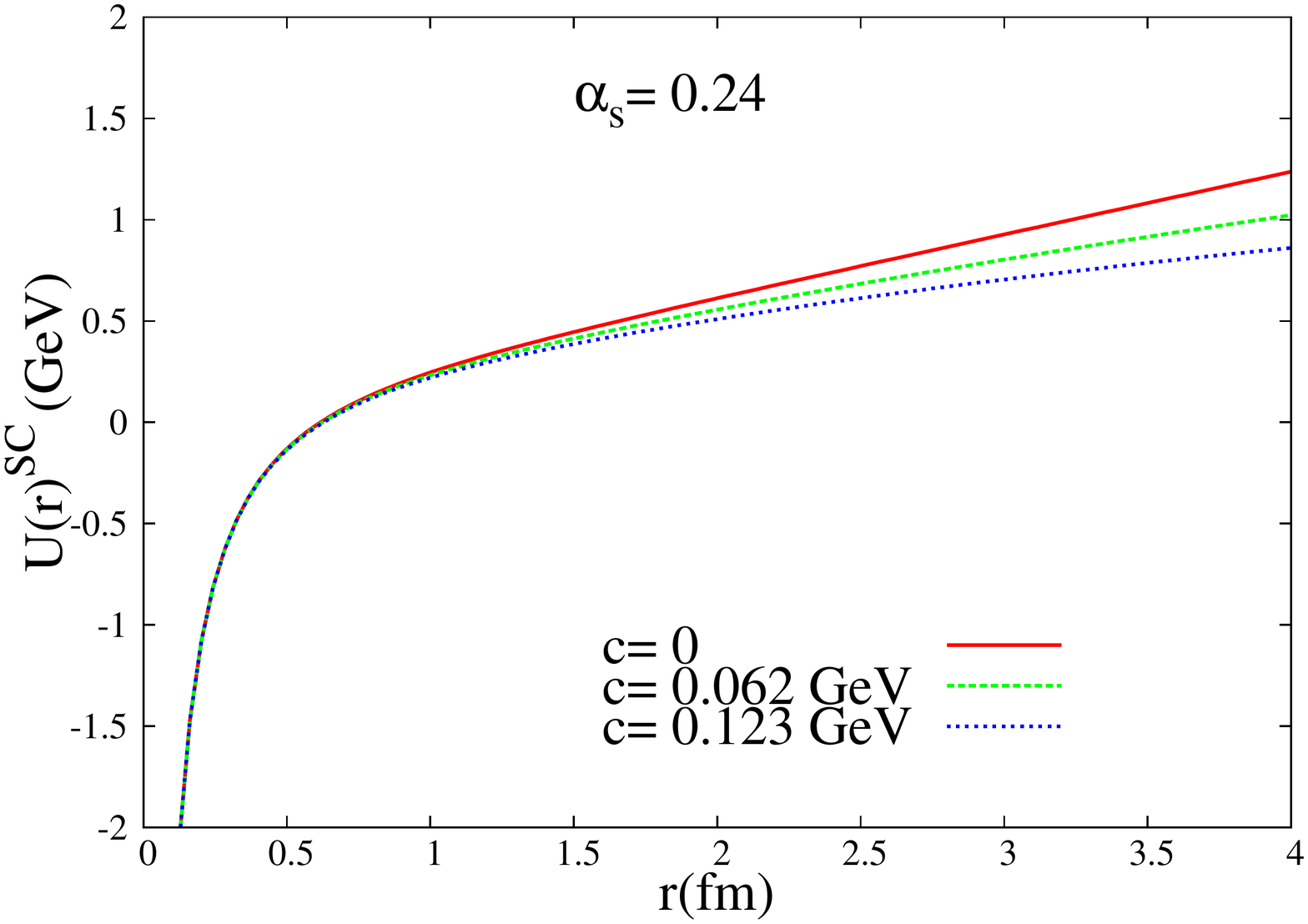}
\includegraphics[width=5cm]{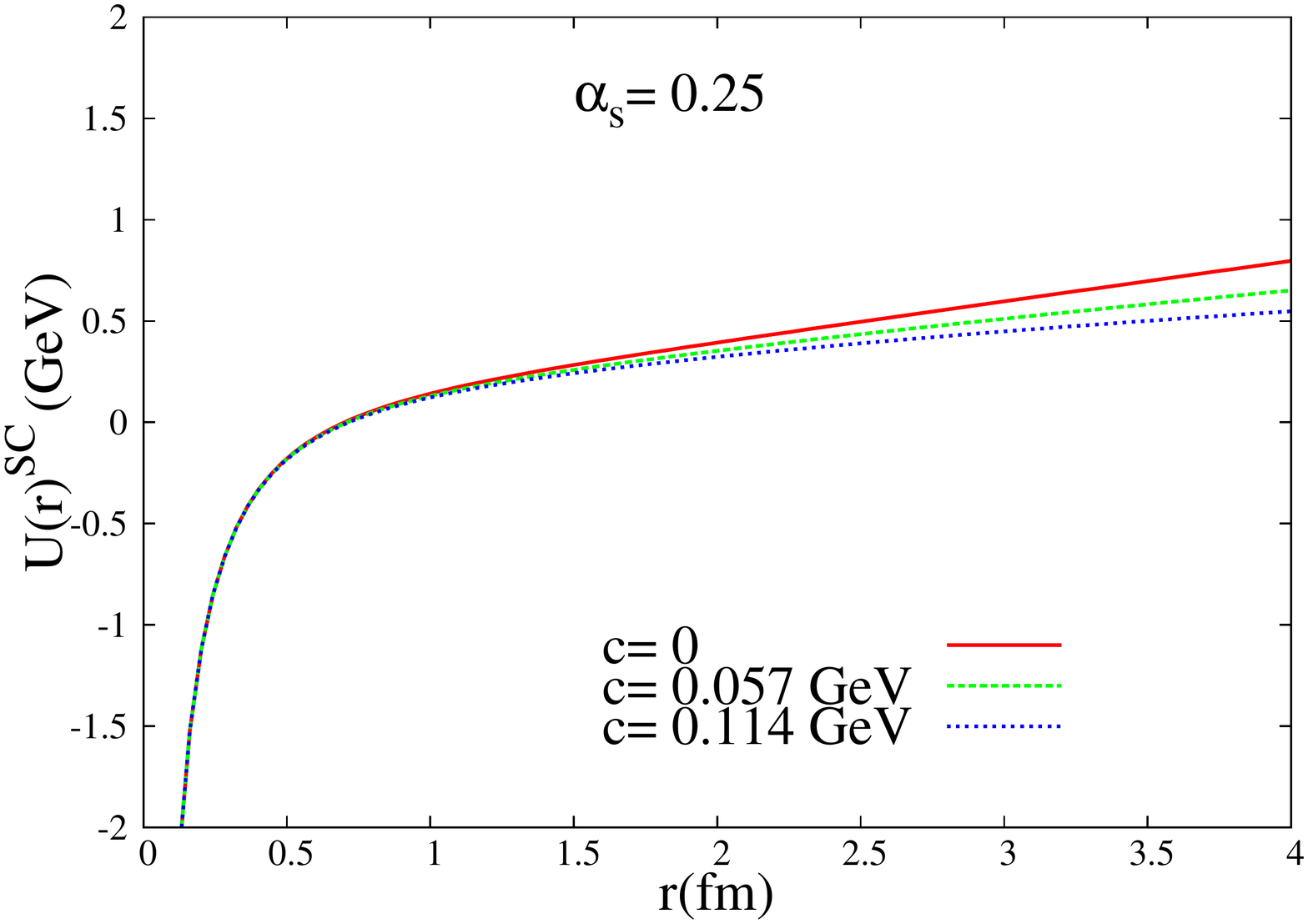}
\caption{(Color online) The quark confining potential with the influence of dynamical quarks with different values of $c$ for $\alpha_s= 0.25$ and $\alpha_s= 0.24$ coupling in $SU(3)$ Dual QCD vacuum.}
\label{Figure3}
\end{figure}

Furthermore, extending the role of dynamical quarks on quark confinement, the action and the associated gluon propagator for $SU(3)$ Dual QCD vacuum has been modified by introducing the infrared cutoff ($a$) \cite{Sasaki} expressed in the following form, 
$$
S_{SU(3)}^{(d)}=\int d^4x \biggl[{j_{\mu}^3}\biggl(-\frac{1}{2}\frac{g^{\mu\nu}}{\partial^{2}+m_{B}^{2}}-\frac{1}{2}\frac{n^{2}}{(n\cdot\partial)^{2}+a^2}\biggl(\frac{m_{B}^{2}}{\partial^{2}+m_{B}^{2}}\biggr) \biggl(g^{\mu\nu}-\frac{n^{\mu}n^{\nu}}{n^{2}}\biggr)\biggr){ j_{\nu}^3}
$$
$$
+{ j}^{8}_{\mu}\biggl(-\frac{1}{2}\frac{g^{\mu\nu}}{\partial^{2}+m_{B}^{'\,2}}-\frac{1}{2}\frac{n^{2}}{(n.\partial)^{2}+a^2}\biggl(\frac{m_{B}^{'\,2}}{\partial^{2}+m_{B}^{'\,2}}\biggr)\biggl(g^{\mu\nu}-\frac{n^{\mu}n^{\nu}}{n^{2}}\biggr)\biggr){j^{8}_{\nu}}\biggr],
$$
and,
\begin{equation}
D^{\mu\nu}_{3}= -\frac{1}{2}\frac{g^{\mu\nu}}{\partial^{2}+m_{B}^{2}}-\frac{1}{2}\frac{n^{2}}{(n\cdot\partial)^{2}+a^{2}}\biggl(\frac{m_{B}^{2}}{\partial^{2}+m_{B}^{2}}\biggr)\biggl(g^{\mu\nu}-\frac{n^{\mu}n^{\nu}}{n^{2}}\biggr),
\label{3.12}
\end{equation}
\begin{equation}
D^{'\,\mu\nu}_{8}= -\frac{1}{2}\frac{g^{\mu\nu}}{\partial^{2}+m_{B}^{'\,2}}-\frac{1}{2}\frac{n^{2}}{(n\cdot\partial)^{2}+a^{2}}\biggl(\frac{m^{'\,2}_{B}}{\partial^{2}+m^{'\,2}_{B}}\biggr)\biggl(g^{\mu\nu}-\frac{n^{\mu}n^{\nu}}{n^{2}}\biggr).
\label{3.13}
\end{equation}
The modified quark confining potential with dynamical quarks under the property of color reflection invariance in the $SU(3)$ Dual QCD vacuum is obtained in the following form,
\begin{equation}
U^{SC}_{SU(3)}(r) = -\frac{{\bf Q}^{2}}{4\pi}\frac{e^{-{\bar m_{B}}r}}{r}+ \frac{{\bf  Q}^{2}{\bar m_{B}}^{2}}{8\pi} \frac{1-e^{-c {\bar m_B} r}}{c {\bar m_B}} \ln\biggl(1+ \frac{(\kappa_{QCD}^{(d)})^2}{1-c^2}\biggr),
\label{3.14}
\end{equation}
where, $c$ is a dimensionless parameter given by, $c = a/\bar m_B$. The resulting plot for the above potential with different values of $c$ are depicted in Fig.~\ref{Figure3} for $\alpha_s= 0.25$ and $\alpha_s= 0.24$ coupling respectively. 

The short range Yukawa part is not modified by the color screening effect and the result obtained for $c=0$ corresponds to the quark confining potential given by ``Eq.(\ref{3.10})''. The influence of light dynamical quarks create vacuum polarization and the linear part of the quark confining potential appears to be screened and such effect gets dominant as $c$ gets larger.

\begin{figure}[ht]
\begin{center}
\includegraphics[width=5cm]{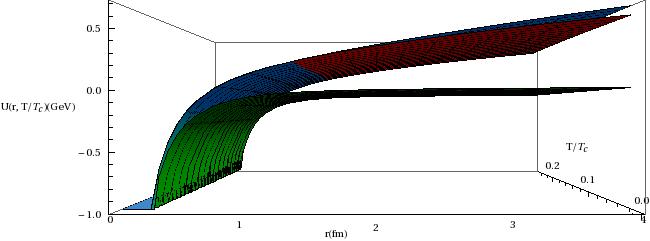}
\includegraphics[width=5cm]{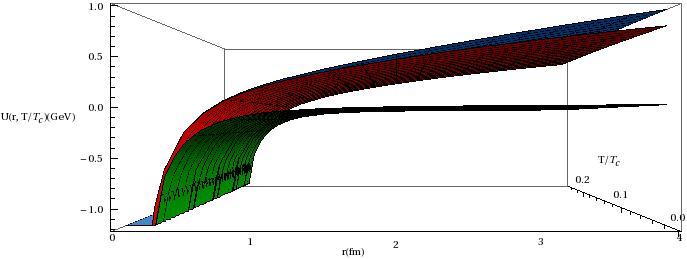}
\includegraphics[width=5cm]{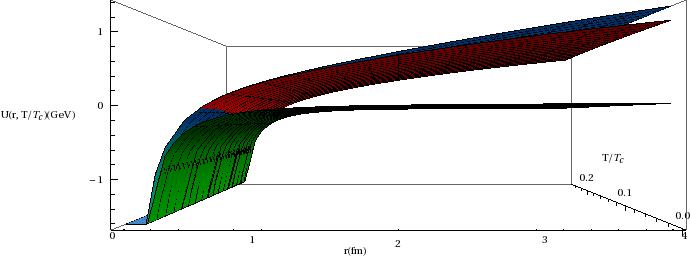}
\caption{(Color online) The temperature dependent quark potential including
color screening effects for several values of $c$ with $\alpha_s=0.25$, $\alpha_s=0.24$ and $\alpha_s=0.23$ coupling in $SU(3)$ Dual QCD vacuum.
}
\label{Figure4} 
\end{center}
\end{figure}

It is clear that, for $r(c\bar m_B) \ll 1$, the confinement potential shows its linear behavior as given by,
\begin{equation}
U_c^{SC}= \frac{\bf Q^2}{8\pi} \bar m_B^2 \ln(1+ \frac{(k_{QCD}^{(d)})^2 }{1-c^2})r.
\label{3.15}
\end{equation} 
On the otherhand, for $r(c\bar m_B) \gg 1$, the confinement potential gets saturated as,
\begin{equation}
U_c^{SC}= \frac{\bf Q^2}{8\pi} \frac{\bar m_B}{c} \ln(1+ \frac{(k_{QCD}^{(d)})^2}{1-c^2}),
\label{3.16}
\end{equation} 
which indicates the dominance of screening at large distances.
The $SU(3)$ Dual QCD based on the magnetic symmetry has its own merits to explain the typical properties of QCD in non-perturbative sector. However, such non-perturbative features are expected to get largely modified in the high temperature region and have important bearings
on the QGP/phase structure of QCD. Let us, therefore, analyse the thermodynamics of $SU(3)$ Dual QCD mainly to study the change in the properties of the QCD vacuum with temperature especially in terms of the QCD monopole condensation and deconfinement phase transition. Using the approach discussed above, we have extended our study to the thermalization and modify the $SU(3)$ Dual QCD quark confining potential in the following way,
$$
U_{SU(3)}(r, T) = -\frac{{\bf Q}^{2}}{4\pi}\biggl[\frac{exp(-\bar m_{B}^{(T)}r)}{r}+\frac{1}{2}(\bar m_{B}^{(T)})^2\biggl(\frac{1-exp(-c\bar m^{(0)}_{B}r)}{c\bar m_{B}^{(0)}}\biggr)
$$
\begin{equation}
 \ln\biggl(\frac{(\bar m_{B}^{(0)})^2(1+(\kappa_{QCD}^{(d)})^2)-(\frac{4\pi\alpha_s+\lambda}{\lambda})T^{2}[\pi\alpha_{s}^{-1}+\frac{3}{2}\lambda\alpha_{s}^{-2}]}{(\bar m_{B}^{(0)})^2-\pi\alpha_{s}^{-1}(\frac{4\pi\alpha_s+\lambda}{\lambda})T^{2}}\biggr)\biggr],
\label{3.17}
\end{equation}
which shows  a finite thermal contribution to the large scale linear confining part of the potential and has its important implications on deconfinement phase transition in QCD. In an identical way, the same analysis may be extended to the case involving the dynamical quarks along with the mean field approximation, we get the associated confining potential in presence of dynamical quarks as given below,
$$
U^{SC}_{SU(3)}(r, T)= -\frac{{\bf Q}^{2}}{4\pi}\biggl[\frac{exp(-\bar m_{B}^{(T)}r)}{r}+\frac{1}{2}(\bar m_{B}^{(T)})^2\biggl(\frac{1-exp(-c\bar m^{(0)}_{B}r)}{c\bar m_{B}^{(0)}}\biggr)
$$
\begin{equation}
 \ln\biggl(\frac{(\bar m_{B}^{(0)})^2(1+(\kappa_{QCD}^{(d)})^2-c^2)-(\frac{4\pi\alpha_s+\lambda}{\lambda})T^{2}[\pi\alpha_{s}^{-1}+\frac{3}{2}\lambda\alpha_{s}^{-2}]}{(\bar m_{B}^{(0)})^2(1-c^2)-\pi\alpha_{s}^{-1}(\frac{4\pi\alpha_s+\lambda}{\lambda})T^{2}}\biggr)\biggr].
\label{3.18}
\end{equation}
In view of the known thermal evolution of the glueballs masses in high temperature domain it is imperative to study the change in the properties of the QCD vacuum at high temperatures especially in the light of QCD-monopole condensation which may further be used to discuss for its observational consequences as string breaking at high temperatures extremely important from the point of view of the heavy-ion collision experiments also and demonstrates a considerable reduction in confinement potential and the associated string tension in the vicinity of critical temperature ($T_c$ ). The variation of the temperature dependent quark potential including
color screening effects for several values of $c$ with $\alpha_s=0.25$, $\alpha_s=0.24$ and $\alpha_s=0.23$ coupling in $SU(3)$ Dual QCD vacuum has been depicted in Fig.~\ref{Figure4} which is in agreement with the recent lattice studies \cite{olaf, Yamamoto}. It further indicates towards the flux tube breaking around $T_c$ and generation of quark-pairs which, in turn, leads to an significant flux screening in high temperature domain.
\section{Results and Conclusions}
The present paper mainly explains the phenomenon of quark confinement within the framework of the $SU(3)$ non-Abelian color gauge theory in terms of the dual Meissner effect due to magnetic monopoles. The Abelian gauge fields have been extracted by inducing a magnetic symmetry and thereby describe the dual dynamics of the non-Abelian monopoles. In the normal phase the magnetic symmetry is preserved, whereas in the confined phase the magnetic symmetry is dynamically broken and ensures the dual Meissner effects due to magnetic condensation that confines colored flux in the QCD vacuum. Using the $SU (3)$ Dual QCD Lagrangian along with the Zwanziger formalism, the associated $SU (3)$ non-perturbative dual gluon propagator has been evaluated and used to derive
the quark confining potential. In the presence of light dynamical
quarks, the $SU (3)$ non-perturbative propagator gets modified by introducing an infrared momentum cutoff parameter $a$ and the quark confining potential is screened in the infrared region due to the quark anti-quark pairs creation. The confinement of quark has been shown to be consistent with the flux tube picture and illustrate the importance of linear confining potential. The quark confining potential in the $SU(3)$ Dual QCD vacuum has been obtained for different values of strong coupling and found to be consistent with the phenomenological Cornell potential $V (r) = \alpha/r + br$ \cite{Chung}. The form of the confining potential for a fixed quark and anti-quark separated by a distance $r$ has also been calculated using lattice gauge theory \cite{Bock, Campbell}. At small distances, the origin of the QCD asymptotic freedom has been revealed in the behavior of quark confining potential. The flux tube starts from the quark and ends to the anti-quark and may be thought of as superposition of several flux-tubes. Moreover, there are two length scales that describe the intrinsic shape and characteristics  of the confining flux-tube. One is the scale that is related to the curvature of the flux tube and sets the exponential decay away from the centre of the flux-tube. It measures the coherence of the magnetic monopole condensate and is called the coherence length ($\xi^{(d)}_{QCD}$ ) of the $SU (3)$ Dual QCD vacuum. The other scale describes the near Gaussian behavior close to the centre of the flux tube and is known as the penetration depth ($\lambda_{QCD}^{(d)}$ ) of the $SU (3)$ Dual QCD vacuum. An attractive Yukawa-like potential \cite{Kinar} may lead
to fluctuating flux-tubes having some intrinsic thickness . The $SU (3)$ Dual QCD vacuum behaves like an effective dual superconductor which belongs to the borderline between a type I and type II superconductor with $\kappa_{QCD}^{(d)} \sim 1$. The existence of the color flux tube provides an explanation for the linearly rising potential for two opposite static color sources. The slope of the potential i.e. the string tension defined as the energy density per unit length of the flux tube approaches a constant value as the distance between the quark and the anti-quark increases. With the introduction to dynamical quarks the flux tube breaks due to the creation of quark and anti-quark pair. As a consequence, the study of the flux tube imposed itself as a tool to investigate the origin of the confining potential for QCD. A finite temperature confining potential has also been extracted from the framework of $SU (3)$ Dual QCD formalism. A considerable reduction in confinement potential
and the associated string tension in the vicinity of critical temperature has been observed and is in agreement with the recent lattice studies. 

\begin{acknowledgments}

Garima Punetha, is thankful to University Grant Commission (UGC), New Delhi, India, for the financial assistance under the UGC-RFSMS
research fellowship during the course of the study.

\end{acknowledgments}


\end{document}